\documentclass[12pt]{article}

\usepackage{bm}% bold math
\usepackage{amssymb}
\usepackage{enumerate}

\usepackage{graphicx} 
\begin{document}

\title{On the internal inconsistency of the Wess-Zumino model} 
\author{N.V.Krasnikov
\\INR RAS,   Moscow 117312, Russia
\\and
\\ JINR, Dubna 141980, Russia
}
%\date{October,1997}
\maketitle
\begin{abstract}
  We prove the internal inconsistency of the supersymmetric Wess-Zumino model. Our proof is based on three assumptions. The first assumption is that in
  the full theory the structure of countertemcs coincides with the structure of the counterterms in the perturbation theory.
The second assumption is the positivity of norm states - no ghosts in the spectrum of the model.
The third  assumption is that the canonical commutation relations among generalized coordinates and momenta  are valid in renormalized theory.
  The obtained results mean that there  are negative norm states in the spectrum of the WZ model. 

\end{abstract}

\newpage

\section{Introduction}

In quantum field theory
renormalizable models
%with single coupling connstant are 
are divided into asymptotically free and asympltotically non free models. Two famous examples  of asymptotically free and
asymptotically non free models are QCD (quantum chromodynamics) and  QED (quantum electrodynamics). In QCD  the effective
coupling constant decreases at small distances therefore  we can use the perturbation theory in the ultraviolet region. The theory looks
self-consistent at least at small distances. While in QED the effective coupling constant increases at small distances
and the perturbation theory is not applicable at small distances. Moreover the ``naive'' use of the perturbation theory leads to the
appearance of the famous Landau pole singularity with negative norm state \cite{bog1}.
% So there is evidence
%  but not a proof \cite{bog1}   that QED is
%  not self-consistent local quantum field theory at small distances.
There is convincing evidence that the four-dimensional
 $\phi^4$ model does not exist as mathematically consistent quantum field theory \cite{phifour1,phifour2,phifour3}.
 All renormalizable  models without nonabelian gauge fields are
non asymptotically free models and probably such models are not self-consistent at small distances.
Therefore it would be very interesting and important to  investigate the problem of consistency or nonconsistency of 
non asymptotically free  renormalizable field theories. In refs. \cite{KRASNIKOV, NIKOLAI} some evidence but not proof has been presented that the
  WZ (Wess-Zumino)  model \cite{WZ1} - \cite{WZ5} is not self-consistent local field theory. See, however  refs.\cite{CRITICS1, CRITICS2}.
In ref. \cite{NUOVOCIMENTO} we   proved that the regime of fixed point or finite wave function renormalization
is not realized in the WZ model, see also \cite{SEEALSO}. 

The aim of this paper is more careful investigation  of the (non)consistency problem in non asymptotically free models
on the example of the WZ     model. We prove that the WZ model is not self-consistent local field theory provided
three  conditions are valid. The first condition is that the structure of the counterterms in the WZ  model coincides with the structure of
  the counterterms in the perturbation theory. The second condition is that commutation relations among generalized coordinates and
momenta  
  are valid in the renormalized field theory. The third condition is that the spectrum of the WZ model does not contain  negative norm states.

  The organization of the paper is the following.  The next section 
contains  short description  of the WZ model. In the third section we describe commutation relations
between the generalized coordinates and the momenta in quantum field theory. In section 4 we derive and analyze Schwinger's equations
for the WZ model. In sections 5 and 6  we  consider  the case of infinite and finite wave function renormalization correspondingly.
    Section 7 contains concluding remarks.

  \section{WZ  model}

%\subsection{Bounds for the GL function in the Wess-Zumino model}

The WZ   model \cite{WZ1} - \cite{WZ5}   describes the interaction of
the scalar and Majorana fields. 
In the superspace    $x^M  = (x^{\mu}, \theta_{\alpha}, \bar{\theta}_{\dot{\alpha}})$ the Lagrangian of the model  has the form
\begin{equation}
  L = L_0  + \int W d^2\theta   + \int W^* d^2{\bar{\theta}}    \,,
  \label{WZ0}
  \end{equation}
where
\begin{equation}
  L_0 = \int \sum_{k=1}^N   \phi_k^*(x,\theta, \bar{\theta})\phi_k(x,\theta, \bar{\theta})   d^2\theta d^2\bar{\theta} \,,
\label{WZ1}
\end{equation}
\begin{equation}
  W =   \sum_{i,j,k}[\frac{g_{ijk}}{3}\phi_i(x, \theta) \phi_j(x, \theta) \phi_k(x, \theta)
    + \frac{m_{ij}}{2}\phi_i(x,\theta)\phi_j(x,\theta)] \,,
    \label{WZ2}
\end{equation}
\begin{equation}
  W^* =   \sum_{i,j,k}[\frac{g^*_{ijk}}{3}\phi^*_i(x, \bar{\theta}) \phi^*_j(x, \bar{\theta}) \phi^*_k(x, \bar{\theta})
    + \frac{m^*_{ij}}{2}\phi^*_i(x,\bar{\theta})\phi^*_j(x,\bar{\theta})] \,,
    \label{WZ2*}
\end{equation}

Here  $\phi_k(x, \theta, \bar{\theta}) = \phi_k(x^{\mu} + i\theta\sigma^{\mu}\bar{\theta}, \theta)$ 
and $\phi_k(x, \theta) = \phi_k(x) + \sqrt{2}\psi_k(x)\theta + \theta \theta F_k(x)$ is chiral scalar superfield.
In terms of component fields $(\phi_k(x), \psi_k(x), F_k(x))$ the Lagrangians
% (\ref{WZ1}, \ref{WZ2})
  have the form 
\begin{equation}
  L_0 =  \sum_{k=1}^N
  [\partial^{\mu}\phi_k \partial_{\mu}\phi^*_k -i\partial_{\mu}\bar{\psi}_k\bar{\sigma}^{\mu}\psi_k + F^*_kF_k] \,,
    \label{WZ2a}
    \end{equation}
 \begin{equation}
   \int W d^2\theta  =   \sum_{i,j,k} g_{ijk}(F_i\phi_j\phi_k - \phi_i\psi_j\psi_k) 
 + \sum_{i,j} m_{ij}(F_i\phi_j -\frac{1}{2}\psi_i\psi_j) \,.
    \label{WZ2b}
 \end{equation}
 \begin{equation}
   \int W^* d^2\bar{\theta}  =   ( \int W d^2\theta)^*       \,.
    \label{WZ2b*}
 \end{equation}

 The WZ Lagrangian (\ref{WZ0}) is invariant under the supersymmetry transformations 
 \begin{equation}
\delta\phi_k =\sqrt{2}\zeta \psi_k \,,
   \label{WZ2ba}
   \end{equation}
 \begin{equation}
\delta\psi_k =i\sqrt{2}\sigma^{\mu}\bar{\zeta}\partial_{\mu}\phi_k +   \sqrt{2}  \zeta F_k \,,
   \label{WZ2bb}
   \end{equation}
  \begin{equation}
\delta F_k =i\sqrt{2}\bar{\zeta}\bar{\sigma}^{\mu}\partial_{\mu}\psi_k  \,.
   \label{WZ2bc}
   \end{equation}
% The WZ model describes  interaction of massive scalar and majorana fields. 

   The WZ model with the  superpotential (\ref{WZ2}) is
renormalizable model  \cite{WZ1} - \cite{WZ5}. Moreover all ultraviolet divergencies can be eliminated by the introduction of the
wave function renormalization counterterm \cite{WZ1}  -  \cite{WZ5}
\begin{equation}
  \Delta L  = \int \sum_{k=1}^N  \Delta  Z_k \phi_k^*(x,\theta, \bar{\theta})\phi_k(x,\theta, \bar{\theta})   d^2\theta d^2\bar{\theta}   
  \,.
  \label{WZ3}
\end{equation}
% and  the corrections to the  superpotential (\ref{WZ2}) are  ultraviolet finite.  
The WZ model has the simplest structure of the counterterms among all renormalizable $d=4$ models with scalar and fermion fields.
For instance, in renormalizable $\phi^4$ model to make the Green's functions ultraviolet finite we have to introduce three different countertems
$\Delta L = \Delta Z_1 \frac{1}{2}\partial^{\mu}\phi\partial_{\mu}\phi - \frac{\delta m^2 }{2}\phi^2  - \Delta Z_2\lambda \phi^4$ .

\subsection{Regularization of the WZ model}

The Feynman diagrams for the WZ model are ultraviolet divergent and to make the model
well defined at least within perturbation theory we have to introduce the regularization.  
  The most convenient regularization
is the supersymmetry invariant regularization. We shall use the
generalization of the Pauli-Villars regularization \cite{PAULI}. For scalar propagator
in the $\phi^4$-model the Pauli-Villars
regularization consists in the replacement
\begin{equation}
  \frac{1}{p^2 - m^2 +i\epsilon} \rightarrow  \frac{1}{p^2 - m^2 +i\epsilon} - \frac{1}{p^2 - M^2 +i\epsilon}
  \label{WZ4}
\end{equation}
 that makes most Feynman diagrams ultraviolet finite. The Pauli-Villars regularization leads to the existence of
negative norm states already at the level of free Lagrangian.  We shall use the modification of the Pauli-Villars
regularization which at the level of the perturbation theory is equivalent to the original Pauli-Villars regularization.
For instance, for $\phi^4$-model we use the following regularized Lagrangian:
\begin{equation}
  L_{reg} = L_{0,reg} + L_{int,reg}  \,,
   \label{WZ5a}
\end{equation}
where
\begin{equation}
L_{0,reg} = \frac{1}{2}[\partial^{\mu}\phi\partial_{\mu}\phi -m^2 \phi^2] +
   \frac{1}{2}[\partial^{\mu}\Phi\partial_{\mu}\Phi - M^2 \Phi^2]   \,, 
   \label{WZ5b}
\end{equation}
\begin{equation}
 L_{int,reg}  = -g(\phi +i\Phi)^4  \,.
\label{WZ5c}
\end{equation}
The  Lagrangian   (\ref{WZ5b})  describes two free
scalar fields $\phi(x)$, $ \Phi(x)$ with masses $m$,  $M$ and positively definite metric.
The interaction Lagrangian (\ref{WZ5c}) is nonhermitean and as a consequence the regularized model is not unitary.
  The propagator for the effective field $   \phi_{eff} = \phi +i\Phi $
coincides with Pauli-Villars propagator (\ref{WZ4}). The generalization of this regularization to the WZ model
is straightforward. Namely we introduce N additional chiral superfields $\Phi_k(x, \theta) $ with masses $M_k $.
The regularized  Lagrangian (\ref{WZ1}) takes the form
\begin{equation}
    L_{0,reg } =  L_0 +  \int \sum_{k=1}^N   \Phi_k^*(x,\theta, \bar{\theta})\Phi_k(x,\theta, \bar{\theta})   d^2\theta d^2\bar{\theta}
+( \int \sum_{k} M_{k}\frac{\Phi_k(x,\theta)\Phi_k(x,\theta)      }{2}     d^2\theta + h.c.)  \,.
\label{WZ5cd}
\end{equation}
  The regularization  of
the interaction Lagrangian
\begin{equation}
  W_{int} = \int \sum_{l,j,k}[g_{ljk}\phi_l(x, \theta) \phi_j(x, \theta) \phi_k(x, \theta) d^2\theta \,
    \label{WZ5d}
 \end{equation}   
    and its hermitean conjugate $W^*_{int}$ is the following:
  \begin{equation}
    W_{int} \rightarrow  \int \sum_{l,j,k}[g_{ljk}(\phi_l(x, \theta) + i\Phi_l(x, \theta))
      (\phi_j(x, \theta) + i\Phi_j(x, \theta))
      (\phi_k(x, \theta) + i \Phi_k(x, \theta)) d^2\theta \,,
\label{WZ6a}
\end{equation}
  \begin{equation}
    W^*_{int} \rightarrow  \int \sum_{l,j,k}[g^*_{ljk}(\phi^*_l(x, \bar{\theta}) + i\Phi^*_l(x, \bar{\theta}))
      (\phi^*_j(x, \bar{\theta}) + i\Phi^*_j(x, \bar{\theta}))
      (\phi^*_k(x, \bar{\theta}) + i \Phi^*_k(x, \bar{\theta}))  d^2{\bar{\theta}}  \,.
\label{WZ6b}
  \end{equation}
  One can find that for modified superpotentials  (\ref{WZ6a}, \ref{WZ6b}) all Feynman diagrams are ultraviolet finite.
  However the interaction Lagrangian is not hermitean as in the case of the $\phi^4$-model. So we find that it is possible
    to reformulate Pauli-Villars regularization
  in such a way that free Lagrangian has positive metric while the interaction Lagrangian is not hermitean and as a consequence the property of the  unitarity
  is lost.

  Another possible regularization  consists in the
  nonlocal generalization  of the vertex $g_{ijk}$ in formula (\ref{WZ5d}), namely
  \begin{equation}
    g_{ijk} \rightarrow  g_{ijk}\exp(-\frac{\vec{p}_i^2}{\Lambda^2} -  \frac{\vec{p}_j^2}{\Lambda^2} -
    \frac{\vec{p}_k^2}{\Lambda^2}    ) \,.
    \label{WZ7}
  \end{equation}
%  In the limit $\Lambda \rightarrow \infty$ of the regularization removing the local limit $g_{ijk}$ is reproduced.

  \section{Canonical quantization}

  In classical  mechanics  with the Lagrangian $L(q_k, \dot{q}_k )$   $(   ~k = 1,2,...N   ) $
    canonical momenta are defined as $p_k = \frac{\partial L(q_k, \dot{q}_k)}{d{\dot{q_k}}}$ and canonical quantization consists in the
    replacement of classical functions $q_k(t)$, $p_k(t)$ to  the  operators $\hat{q}_k(t)$ and $\hat{p}_k(t)$ acting in Hilbert space.
    The  commutation relations\footnote{In this paper we use the natural units with $\frac{h}{2\pi} \equiv 1 $}
      \begin{equation}
        [\hat{q}_k(t), \hat{q}_l(t)]  = [\hat {p}_k(t), \hat {p}_l(t) = 0  \,
\label{Can1}
      \end{equation}
      \begin{equation}
       i [\hat {p}_k(t),\hat{q}_l(t)] =  \delta_{kl} \,
        \label{Can2}
      \end{equation}
      are postulated.
      In the field theory with the complex scalar fields $\phi_k(x)$ and with the Lagrangian
      $L = \sum_{k=1}^{N}\partial_{\mu}\phi_k(x)\partial^{\mu}\phi^*_k(x) - V(\phi_k(x), \phi^*_k(x))$ the density of the
      momenta is $ \pi_k(t, \vec{x}) = \frac{\partial \phi_k(t, \vec{x})}{dt}$ 
      and the commutation relations are \cite{bog1}
     \begin{equation}
  [\phi_k(t,\vec{x}), ~\phi_l(t,\vec{y})] = [\phi^*_k(t,\vec{x}), ~\phi_l(t,\vec{y})] = 0 \,, 
        \label{Can3a}
        \end{equation}
        \begin{equation}
[\frac{\partial_k\phi_k(t,\vec{x})}{dt}, ~\phi_l(t,\vec{y})]  = 0 \,, 
        \label{Can3b}
        \end{equation}
  \begin{equation}
[\frac{\partial_k\phi_k(t,\vec{x})}{dt}, ~\phi_l^*(t,\vec{y})]  = 0   ~at ~k \neq l \,,
        \label{Can3c}
        \end{equation}
  \begin{equation}
  i [\frac{\partial_k\phi_k(t,\vec{x})}{dt}, ~\phi_k^*(t,\vec{y})]  = \delta(\vec{x} - \vec{y})    \,.
        \label{Can3d}
  \end{equation}
  
  The  renormalized field  $  \phi_{kr\Lambda}(x) $ is
  proportional to the bare field $ \phi_k(x)$, namely \cite{bog1}  $\phi_{kr\Lambda}(x)  =  Z^{-1/2}(\Lambda,...) \phi_k(x)$. Here $\Lambda$ is ultraviolet
  cutoff. Note  that  the limit of the regularization removing $\Lambda \rightarrow \infty $ exists at least in the perturbation theory.
  We shall assume that the limit $\Lambda \rightarrow \infty $ exists irrespective of the perturbation theory, in other words we assume that
  the $lim_{\Lambda \rightarrow \infty} ~\phi_{kr\Lambda}(x) = \phi_{kr}(x)$ exists.
  As a consequence   the commutation relations
  (\ref{Can3a}~-~\ref{Can3c}) are valid for  renormalized fields. The commutation relation (\ref{Can3d}) takes the form
 \begin{equation}
i[\frac{\partial_k\phi_{kr}(t,\vec{x})}{dt}, ~\phi_{kr}^*(t,\vec{y})]  = \frac{1}{Z_k(\infty,...)} \delta(\vec{x} - \vec{y})    \,.
        \label{Can3dr}
 \end{equation}
 We assume  that  for all states $|n>$ the metric is positive, i.e. $<n|n> > 0$.
 As a consequence of this assumption  the 
 KL (Kallen -Lehmann) representation \cite{KL1, KL2}  for two point vacuum  commutator $<0|[\phi_{kr}(x), \phi^*_{lr}(y)]|0>$
 reads
 \begin{equation}
   <0|[\phi_{kr}(x), \phi^*_{lr}(y)]|0> = \frac{1}{i}\int_{0}^{\infty} D(x-y,t)\rho_{kl}(t) dt \,,
     \label{Can4a}
 \end{equation}
 where
 \begin{equation}
   D(x-y, m^2) = \frac{i}{(2\pi)^3}\int  \exp{[-ik(x-y)]} \epsilon(k^0)\delta(k^2-m^2)d^4k \,,
   \label{Can4d}
   \end{equation}
\begin{equation}
  \rho_{kl}(t) = c^2_k\delta_{kl}\delta(t- m^2_k) + \Delta\rho_{kl}(t) \,
  \label{Can4d1}
\end{equation}
and $c^2_k  > 0$, $ \Delta\rho_{kk}(t) \geq  0$. The first term in the
formula (\ref{Can4d1}) describes one particle state contribution while
the second term  describes many particles contribution.
 As a consequence of the commutation relations (\ref{Can3c}, \ref{Can3dr}) one can find that
 \begin{equation}
   \int_{0}^{\infty}\rho_{kl}(t)dt = 0 ~at~ k \neq l \,,
   \label{Can5d}
\end{equation}
\begin{equation}
   \int_{0}^{\infty}\rho_{kk}(t)dt =    \frac{1}{Z_k(\infty,...)} \,.
   \label{Can5dd}
   \end{equation}
   For  often used  normalization condition $c^2_k = 1$ well known inequality \cite{KL1, KL2} 
   \begin{equation}
 0 \leq Z(\infty,...) = [1 +  \int_{0}^{\infty}\Delta \rho_{kk}(t)dt]^{-1} \leq 1 \,,
   \label{Can5e}
   \end{equation}
   is valid.
   For the propagator $\frac{1}{i}\bar{D}_c(x-y) = <0|T(\phi_r(x), \phi_r^*(y))|0> $ of the
   renormalized scalar field $\phi_r(x)$ the KL representation takes the form
   \begin{equation}
     \bar{D}_c(x) = \frac{1}{(2\pi)^4}\int \exp({-ikx})D_c(k^2) \,,
     \label{Can5e1}
   \end{equation}
   \begin{equation}
     D_c(k^2) = \frac{1}{m^2_{0} - k^2 - i\epsilon} + \int_{4m^2_o}^{\infty}   dt \frac{\Delta\rho(t)}{t - k^2 -i\epsilon} \,,
     \label{Can5e2}
   \end{equation}
   where $  \Delta\rho(t) \geq 0 $.
   As a consequence of the commutation relation (\Ref{Can3dr}) and the KL representation (\Ref{Can5e2}) one can  find that
  only two possibilities are possible.
    Namely, we can have  finite
    wave function renormalization  $Z(\infty, ...) \neq 0$ or infinite wave function renormalization  with   $Z(\infty, ...) = 0$.
    The possibility with $Z(\infty, ...) = \pm \infty$ is excluded due to assumed nonnegativity of $\Delta \rho_{kk}(t)$.  The infinite
     $Z(\infty, ...)  $ 
    means the negativity of   $\Delta \rho_{kk}(t_0) < 0 $ at some $t = t_0$, i.e. the existence of negative norm states in the spectrum.
    For    $Z(\infty, ...) \neq 0 $    the ultraviolet behaviour of the propagator (\Ref{Can5e2}) coincides
    up to some factor with free massless propagator $\frac{1}{-k^2- i\epsilon}$
   while for infinite renormalization with  $Z(\infty, ...) =  0$ the propagator decreases more slowly than free propagator, i.e.
   $k^2 D_c(k^2) \rightarrow \infty $ at $k^2 \rightarrow \infty $.

   It should be noted that for the WZ model   in the leading order of the perturbation
   theory $Z(\Lambda, ..) = 1 - ag^2\ln(\Lambda) + o(g^2)$  $ ( a >0)$ 
   and $Z \rightarrow - \infty$
   at $\Lambda \rightarrow  \infty  $. The manifestation of this fact is the existence of Landau pole singularity for the scalar field propagator
   in leading log approximation.

 \section{Schwinger's equations}

 The Schwinger's equations \cite{Schwinger} are mostly conveniently obtained in the functional integral formalism by means of the identity
 \begin{equation}
   \int\prod_{k} d[\phi_k]
   \prod_{i}
   \frac{\delta}{\delta \bar{\phi}_i(x_i)}[R(\phi_k) \exp(iS(\Phi_k)] = 0 \,.
       \label{Sch1}
 \end{equation}
 Here $\phi_k$ denotes the multiplet $(\phi_k, F_k, \psi_k)$,
 the action $    S(\Phi_k) =      \int d^4x L +\int J_k\phi_{k} d^4x $ and
 $R[\phi_k] $ an arbitrary function; $\bar{\phi_i}$ may be taken to be any of the fields  $\phi_k$ $ F_k $ and $ \psi_k $.
The crusial point is that for renormalized Lagrangian (\ref{WZ6a}, \ref{WZ6b}) and for finite regularization $\Lambda$ 
we can derive Schwinger's equations for any $\Lambda $. It should be stressed that we assume finite limit for
$Z(\Lambda,...)$ at $\Lambda \rightarrow \infty$.
% The finiteness of the $Z(\infty,...)$
This assumption is not valid in the perturbation theory where at one loop level $ Z = 1 - ag^2_r \ln(\frac{\Lambda}{\mu})$, $a  > 0 $.

\section{The infinite renormalization $Z(\infty, ...) = 0   $   }

As an example consider  the  WZ model with three chiral superfields $\phi_k(x, \theta)   =  \phi_{k}(x)  +
\sqrt{2}\psi_k(x)\theta + F_k(x)\theta\theta $
 ($k = 1,2,3$) and with the superpotential
\begin{equation}
  W   = W_2 + W_3 \,,
  \label{infinite1a}
  \end{equation}
\begin{equation}
  W_2      =    \frac{m}{2}(\phi^2_1(x,\theta) + \phi^2_2(x, \theta) + \phi^2_3(x, \theta)) \,,
  \label{infinite01}
\end{equation}
\begin{equation}
  W_3 = g\phi_1(x, \theta)\phi_2(x, \theta)\phi_3(x,\theta)    \,.
  \label{infinite1}
\end{equation}
The kinetic term  has standard form
\begin{equation}
  L_{0} =   \int d^2\theta d^2{\bar{\theta}}   \sum_{k=1}^3 \phi^*_k(x, \theta, \bar{\theta})\phi_k(x, \theta, \bar{\theta})  \,,
  \label{infinite1b}
  \end{equation}
The Lagrangian  with the superpotential
(\ref{infinite1a}, \ref{infinite01}, \ref{infinite1}) is invariant under the
  discrete transformations $\phi_k \rightarrow \phi_l$, $\phi_l \rightarrow \phi_k$.
  As a consequence the counterterm
  \begin{equation}
    \Delta L =   \int d^2\theta d^2{\bar{\theta}}
    \sum_{k=1}^{3}( Z(\Lambda, g_r, \mu,  m_r) - 1)\phi^{*}_k(x, \theta, \bar{\theta}) \phi_k(x, \theta, \bar{\theta})
\label{infinite2}
  \end{equation}
  makes the Green's functions ultraviolet finite in each order of the perturbation theory. Here we use
  the ultraviolet supersymmetric regularization  with three additional chiral superfields
  $\Phi_k(x, \theta)$  with additional superpotential $\Delta W = \sum_{k=1}^3 \frac{\Lambda}{2}\Phi_k(x,\theta)\Phi_k(x, \theta) $
  and kinetic term  $L_{0\Phi} =        \int d^2\theta d^2{\bar{\theta}}   \sum_{k=1}^3 \Phi^*_k(x, \theta, \bar{\theta})\Phi_k(x, \theta, \bar{\theta})$.
  The chiral superfields  $\Phi_k(x, \theta)$ describe massive scalar and Majorana fields with a mass $\Lambda$. The regularization
  with additional chiral superfields  $\Phi_k(x, \theta)$ consists in the replacements $\phi_k(x, \theta) \rightarrow
  \phi_k(x, \theta) + i \Phi_k(x, \theta),   ~~ \phi_k^{*}(x, \bar{\theta}) \rightarrow \phi_k^{*}(x, \bar{\theta})
  + i \Phi_k^{*}(x, \bar{\theta}) $ for the superpotential $W_3 = g\phi_1(x, \theta)\phi_2(x, \theta)\phi_3(x,\theta)$
  and $W_3^* = g\phi^*_1(x, \bar{\theta})\phi_2^*, \bar{\theta})\phi_3^*,\bar{\theta})$.
  The regularizarion with the introduction of additional massive chiral superfields $\Phi_k(x, \theta)$
  preserves the supersymmetry and at the level of the absence of the interaction $g = 0$ there are no ghost states in
  the spectrum. However the  interaction Lagrangian is not hermitean that leads to the violation  of the unitarity for
  regularized Lagrangian.
  %It is also possible to introduce nonlocal regularization of the
  %three point local vertex $g$, namely
  %\begin{equation}
  %  g \rightarrow g \exp(-\frac{\vec{p}_1^2}{\Lambda^2} -\frac{\vec{p}_2^2}{\Lambda^2} -\frac{\vec{p}_3^2}{\Lambda^2}    ) \,.
  %      \label{infinite3}
  %\end{equation}
 % Both regularizations preserve the hamiltonian interpretation of the theory. For the renormalized Lagrangian
  %  $L_{0\phi} +L_{0\Phi} + \Delta L  +   \Delta W + W + W^* $
  As a consequence of the Schwinger's equations we find  that
 % for scalar field $\phi(x)$
  \begin{equation}
    - Z(\Lambda,...) <0|T(\phi^*_{r1}(x, \Lambda), F^*_{r1}(y, \Lambda)|0> = m_r<0|T(\phi^{*}_{r1}(x, \Lambda), \phi_{r1}(y, \Lambda)|0> +
    %    g_r <0|T(\phi^{*}_{r1}(x, \Lambda), (\phi_{r2}(y, \Lambda)\phi_{r3}(y, \Lambda))_{reg})|0>  \,,
\label{Infinite4}
  \end{equation}
$$  g_r <0|T(\phi^{*}_{r1}(x, \Lambda), (\phi_{r2}(y, \Lambda)\phi_{r3}(y, \Lambda))_{reg})|0>, $$
  
 where $\phi_{rk}(x, \Lambda) = Z^{-1/2}( \Lambda, ...)\phi_k(x)$, $m_r =  Z(\Lambda, ...)m$, 
 $g_r =  Z^{3/2}(\Lambda, ...)g$, $(\phi_{r2}(y, \Lambda)\phi_{r3}(y, \Lambda))_{reg} =
 (\phi_{r2}(y) +i\Phi_{r2}(y)) (\phi_{r3}(y)+ i\Phi_{r3}(y) )   $.

 We assume that 
 the  limit
  $\Lambda \rightarrow \infty$ exists for 
 the Green's functions of  the renormalized fields $\phi_{rk}(x, \Lambda)$, namely we assume that
 \begin{equation}
   ( \phi_{rk}(x, \Lambda),  \psi_{rk}(x, \Lambda), F_{rk}(x, \Lambda)  \rightarrow  ( \phi_{rk}(x),  \psi_{rk}(x) , F_{rk}(x))
   ~ at~ \Lambda \rightarrow \infty \,
   \label{ASSUMP}
   \end{equation}
 As a consequence of the assumption  (\ref{ASSUMP}) 
% that   $Z(\Lambda, g_r, \mu, m_r  ) \rightarrow 0$ at $\Lambda \rightarrow \infty$
    the equation (\ref{Infinite4}) takes the form
 \begin{equation}
   m_r<0|T(\phi^{*}_{r1}(x), \phi_{r1}(y)|0> +  g_r <0|T(\phi^{*}_{r1}(x), (\phi_{r2}(y)\phi_{r3}(y)))|0> = 0 \,.
\label{Infinite4a}
 \end{equation}
 The analogous relation for the spectral densities of the KL representations  for the propagators reads 
 \begin{equation}
m_r\rho_{\phi_{r1}^*, \phi_{r1}}(t) + g_r\rho_{\phi_{r1}^*, \phi_{r 2}\phi_{r 3}}(t)  = 0 \,.
\label{Infinite7a}
 \end{equation}
%The second very important step is that
 %For renormalized scalar fields
 %  $\phi_{rk} = Z^{-1/2}(\Lambda,...)\phi_{k}(x)$   the commutation relations
 %(\ref{Can3a} - \ref{Can3c})  also take place.
 As a consequence of the commutation relations 
 (\ref{Can3a} - \ref{Can3c}) we find that
 \begin{equation}
   [\frac{\partial\phi^{*}_{r1}(t, \vec{x})}{\partial t},  (\phi_{2r}(t, \vec{y})\phi_{3r}(t, \vec{y}))_{reg}] = 0 \,,
   \label{Infinite5}
 \end{equation}
% Earlier we assumed that the limit $\Lambda \rightarrow \infty$ exists for the renormalized fields.
% As a consequence of this assumption the commutation relation (\ref{Infinite5}) is valid also
% in the limit  $\Lambda \rightarrow \infty$
%  for renormalized fields.
For the commutation relation (\ref{Infinite5}) the equation  (\ref{Can5d}) takes the form
 \begin{equation}
 \int_{0}^{\infty}\rho_{\phi^*_{r1}, \phi_{r2}\phi_{r3}} (t)dt = 0 \,.
 \label{Infinite6}
 \end{equation}
 As a consequence of (\ref{Infinite7a}) we find that
\begin{equation}
 \int_{0}^{\infty}\rho_{\phi^*_{r1}, \phi_{r1}} (t)dt = 0 \,.
 \label{Infinite66}
 \end{equation}
%Due to the assumed positivity of the metric for the states  ($<n|n>  ~>~  0$) the spectral density $ \rho_{\phi_{1r}^*, \phi_{1r}}(t) \geq 0 $
% is non negative.
Due to  the relation  (\ref{Infinite66})
and nonegativity      $ \rho_{\phi_{1r}^*, \phi_{1r}}(t) \geq 0 $    of the spectral density  
\begin{equation}
 \rho_{\phi^*_{r1}, \phi_{r1}} (t) = 0 \,.
 \label{Infinite66}
 \end{equation}
The analogous equalities are valid for spectral densities $ \rho_{\phi^*_{r2}, \phi_{r12}} (t) $ and
  $ \rho_{\phi^*_{r3}, \phi_{r3}} (t)$.
% \begin{equation}
%   m_r g_r\rho_{\phi_{r1}^*, \phi_{ 2}\phi_{ 3}}(t) \leq 0 \,
%   \label{Infinite7b}
% \end{equation}
  It means that  the  scalar propagators vanish and the spectrum of the model is empty   that contradicts to the
  expectations from the perturbation theory that the spectrum  contains  the scalar massive state\footnote{According to common lore
    the perturbation theory is valid  for small effective coupling constants.}.

% is non positive.   Moreover the relations (\ref{Infinite7b}) and  (\ref{Infinite6}) lead to zero   spectral density
% $m_rg_r\rho_{\phi_{r1}^*, \phi_{ 2}\phi_{ 3}}(t) = 0 $ and 
% $\rho_{\phi_{r1}^*, \phi_{r1}}(t)  = 0$.
% So the  scalar propagator vanishes  that contradicts to the
%  expectations from the perturbation theory that the spectrum  contains  the scalar massive state.

 Another way to understand the inconsistency of the infinite wave function renormalization  $Z(\infty,...) = 0$ is the following \cite{NIKOLAI}.
  Take the differential operator in (\ref{Sch1}) to be $\frac{\delta}{\delta F_{r1}(x)}\frac{\delta}{\delta F^*_{r1}(x)}$.
In the limit $\Lambda \rightarrow \infty $ of the regularization removing  one can find \cite{NIKOLAI} that
 \begin{equation}
   <0|T(m_r\phi_{r1}(x) + g_r\phi_{r2}(x)\phi_{r3}(x), m_r\phi^*_{r1}(y) + g_r\phi^*_{r2}(y)\phi^*_{r3}(y)|0> = 0 \,.
\label{NIKOLAI1}
 \end{equation}
 Using the equation (\ref{Infinite4a}) one can find that
 \begin{equation}
   <0|T(m_r\phi_{r1}(x), m_r\phi^*_{r1}(y))|0> =
   <0|T( g_r\phi_{r2}(x)\phi_{r3}(x),  g_r\phi^*_{r2}(y)\phi^*_{r3}(y)|0> \,.
   \label{NIKOLAI2}
   \end{equation} 
 Due to the absence of radiative corrections to the superpotential (\ref{WZ2}) $ m_{ij}\phi_i(x, \theta)\phi_j(x, \theta) =
  m_{rij}\phi_{ri}x, \theta)\phi_{rj}(x, \theta) $.
  As a consequence the composite renormalized operator $(\phi_{i}(x, \theta) \phi_{j}(x, \theta))_r = \phi_{ri}(x, \theta)\phi_{rj}(x, \theta) $
  and the same equality $(\phi_{i}(x)\phi_{j}(x))_r = \phi_{ri}(x)\phi_{rj}(x) $ takes place for scalar fields
  $\phi_{ri}(x)$ and $\phi_{rj}(x)$.  Using the KL representation for the equality  (\ref{NIKOLAI2})  one can find that the spectral densities
  satisfy the equation
  \begin{equation}
    m^2_r\rho_{\phi_{r1}, \phi^*_{r1}}(t) = g^2_r\rho_{(\phi_1\phi_2)_r,(\phi^*_1\phi^*_2)_r))} \,
  \label{NIKOLAI3}
  \end{equation}
  and all terms in (\ref{NIKOLAI3}) are well defined within perturbation theory at sufficiently small renormalization coupling
  constant $g_r$. For small $g_r$ we have $ m^2_r\rho_{\phi_{r1}, \phi^*_{r1}}(t) \sim m^2_r\delta(t - m^2_r) $ while the right hand side of the equation
  (\ref{NIKOLAI3}) is equal  to $\frac{g^2_r}{16\pi^2}(1 -\frac{4m^2_r}{t})^{1/2}  + O(g^4_r)  $. So we find the contradiction. The same contradiction
  takes place for the equation (\ref{Infinite7a}).
  
 The limit $Z(\infty, ...) = 0$  is described by the ultralocal
 Lagrangian  
 \begin{equation}
   L_{ultra} = \int d^2\theta [ (g_r \phi_{r1}(x,\theta)\phi_{r2}(x,\theta)\phi_{r3}(x,\theta) +\frac{m_r}{2} \phi_{r1}^2(x,\theta)
     + ...]    + h.c. \,
  \label{Infinite9}
  \end{equation}
without derivatives. The Lagrangian (\ref{Infinite9})
 has been considered  in ref.\cite{NIKOLAI1}. For the Lagrangian (\ref{Infinite9}) all scalar Green's functions
 vanish. It is easy to prove this fact using the continual integral formalism for the Lagrangian
 (\ref{Infinite9}). Really for  the Lagrangian $L_{utra}$ after the integration over the auxiliary  fields $F(x)$, $F^*(x)$
 we find the delta functions
 \begin{equation}
   \int dF_{rk}(x)dF_{rk}^{*}(x) d...exp(iL_{ultra}) \sim \int d... \delta(m_r\phi_{r1}(x) + g_r\phi_{r2}(x)\phi_{r3}(x))...
    \,.
   \label{Infinite9a}
 \end{equation}
 inside of the integral 
 that allows to calculate the continual integral for $L_{ultra}$ exactly \cite{NIKOLAI1}. The spectrum for the Lagrangian
 $L_{ultra} $ is an empty, i.e. it does not contain any physical states.
 
 Note that it is possible also to consider the standard case with single scalar chiral superfield
 $\phi(x, \theta)$ and with the superpotential $W = \frac{m}{2}\phi^2(x, \theta) + \frac{g}{3} \phi^3(x, \theta) $.
 For such potential the equations (\ref{Infinite4a}, \ref{Infinite7a}) are also valid with the replacement
 $g_r\phi_{r2}(y)\phi_{r3}(y) \rightarrow g_r\phi^2_r(y)$.  In the derivation of the sum rule
 $\int_{0}^{\infty}\rho_{\phi^*_{r}, \phi_{r}\phi_{r}} (t)dt = 0 $ we have to use in addition the commutation relation (\ref{Can3dr}).
 As in a previous case we find that the infinite renormalization  $Z(\infty, ...) = 0$    is not realized.

 \section{Finite renormalization $Z(\infty, ...  )  \neq 0$ }
 
 In this section we repeat  the main results of \cite{NUOVOCIMENTO} where
 we   proved that finite renormalization  is
 not realized     in the WZ model  with
 % chiral superfield $\phi(x, \theta)$ and with
 the  superpotential
 $W = \frac{m}{2} \phi^2(x, \theta)  +    \frac{g}{3}\phi^3(x, \theta)$. For finite renormalization  the ultraviolet asymptotics of
 the scalar  propagator
        coincides up to some factor  with free massless scalar propagator.
 %       namely $D_{\phi^*,\phi}(p^2, m_r, \mu, g_r) \rightarrow \frac{1}{- Z(\infty,...) p^2}$ at $p^2 \rightarrow \infty$.
        Due to the relation $g = g_r Z^{-3/2}(\Lambda,...)$ between bare charge $g$ and renormalized charge $g_r$
        finite renormalization corresponds to fixed point for the beta function, $\beta(g_{r, fp}) = 0 $.
      Moreover   the  ultraviolet behaviour of other 
      Green's functions coincides with the behaviour of the massless WZ model at fixed point $g_r = g_{r, fp}$.
      It is easy to prove this fact \cite{NIKOLAI} using the renormalization group.
      Namely the renormalization group  equation in the Weinberg  $-$ t`Hooft scheme \cite{W,H}\footnote{
        In the Weinberg  $-$ t`Hooft scheme  $Z(\Lambda, g_r ,  \mu, m_r)$
      doesn't   depend on the mass $m_r$.} for n-point Green's function $G_n(p_1, ...,p_n)$  can be written in the form
      \begin{equation}
        [\mu\frac{ \partial}{\partial \mu}  + \beta(g_r)\frac{\partial}{\partial g_r} + m_r \gamma_m(g_r) \frac{\partial}{\partial m_r}
          +n\gamma(g_r)]G_n = 0 \,,
          \label{eqrg}
      \end{equation}
      where $\beta(g_r) = 3 g_r \gamma(g_r) $ and $\gamma_m(g_r) = 2 \gamma(g_r)$.
      %   It is well known  that fixed point
    %  $g_{r, fp}$ with   $\beta(g_{r, fp}) = 0$  determines 
    %     the ultraviolet behaviour of the Green's function (\Ref{eqrg}).
          As a consequence of
      the relation   $\beta(g_r) = 3g_r \gamma(g_r)$ between the $\beta$-function and the anomalous dimension $\gamma(g_r)$
      at fixed point $g_{r, fp}$ the anomalous dimension   $\gamma(g_{r, fp}) = 0 $ and the ultraviolet behaviour of the
      scalar propagator coincides up to some normalization
      factor with free scalar propagator.
  We shall use the Schwinger's equations ({\Ref{Sch1}) with $\frac{\delta}{\delta F_r(x)}$.
       As in the previous section we  consider   massless WZ model with three chiral scalar
       superfelds $\phi_k(x, \theta)$ $(k= 1,2,3)$
       and
            with the superpotential (\Ref{infinite1}). Due to the symmetry $\phi_i(x, \theta) \rightarrow \phi_j(x, \theta )$,
       $\phi_j(x, \theta) \rightarrow \phi_i(x, \theta) $ of the superpotential  (\Ref{infinite1}) $ \gamma_{i}(g_r) = \gamma_j(g_r)
       \equiv \gamma(g_r) $ and
         $\beta(g_r) = 3 g_r \gamma (g_r)$.
   As a consequence of the Schwinger's equations we find
        \begin{equation}
          Z(\infty,...) \delta(x) + i  Z^2(\infty, ...) <0|T(F^{*}_{r1}(x), F_{r1}(0)|0> = ig^2_r <0| T(\phi_{r2}(x)\phi_{r3}(
            x), \phi^{*}_{r3}(0)\phi^{*}_{r2}(0)>  \,. 
            \label{masslesseq}
    \end{equation}           
        The analogous equations take place for the scalar fields $\phi_{r2}$, $\phi_{r3}$.
        In momentum space the equation (\ref{masslesseq}) has the form
        \begin{equation}
          Z(\infty,...) + Z^2(\infty,...) D_{F_{r1},F_{r1}^*}(p^2) = g^2_r D_{\phi_{r2}\phi_{r3}, \phi^*_{r3}\phi^*_{r2}}(p^2) \,. 
          \label{massl1}
          \end{equation}
        As a consequence of  the supersymmetric Ward identities \cite{WZ1} -  \cite{WZ5}  
        $D_{F_{r1},F^*_{r1}}(p^2) = p^2 D_{\phi_{r1},\phi^*_{r1}}(p^2)$ the analog of the equation (\ref{massl1})
         for the spectral densities of the KL representations reads
        \begin{equation}
          Z^2(\infty,...) t \rho_{\phi_{r1},\phi^*_{r1}}(t) = g^2_r\rho_{\phi_{r2}\phi_{r3}, \phi^{*}_{r3}\phi^{*}_{r2}}(t) \,.
            \label{KLrho}
        \end{equation}
        Due to assumed finiteness of $ Z^{-1}(\infty,...) = \int^{\infty}_{0} \rho_{\phi_{r1},\phi^*_{r1}}(t) dt$
        we find that $ lim_{t \rightarrow \infty} (t \rho_{\phi_{r1},\phi^*_{r1}}(t))  = 0$.
        The spectral density $ g^{2/3}_r \rho_{\phi_{r1},\phi^*_{r1}}(t)$ has zero anomalous dimension and 
         as a consequence of the commutation relations (\ref{Can3dr})    $ Z(\infty,...) = (\frac{g_r}{g_{r,fp}})^{2/3}$.
        The spectral density $Z(\infty,...)\rho_{\phi_{r1},\phi^*_{r1}}(t)$       can be represented in the form
        \begin{equation}
          Z(\infty,...)\rho_{\phi_{r1},\phi^*_{r1}}(t) = \delta(t) + \frac{1}{t}\Delta \rho(\frac{t}{\Lambda^2}) \,,
          \label{ZREN}
        \end{equation}
        
        where $\Lambda$ satisfies the renormalization group equation
        $ (\mu\frac{ \partial}{\partial \mu}  + \beta(g_r)\frac{\partial}{\partial g_r})\Lambda   =0 $ and
        the  $\Lambda = 0$ corresponds to the  fixed point $g_{r, fp}$ with $\beta(g_{r,fp}) = 0$.
        At fixed point $ Z(\infty,...)\rho_{\phi_{r1},\phi^*_{r1}}(t) = \delta(t)$ and the 
        $  lim_{t \rightarrow \infty} \Delta \rho(\frac{t}{\Lambda^2}) = 0$.
        The right hand side of the equation  (\ref{KLrho}) can be represented in the form
        \begin{equation}
      g^2_r\rho_{\phi_{r2}\phi_{r3}, \phi^{*}_{r3}\phi^{*}_{r2}}(t) = Z(\infty,...)\Phi(t/\Lambda^2)   \,.
          \label{KLrho1}
        \end{equation}
        As it has been mentioned before the case $\Lambda = 0$ or the limit $t \rightarrow \infty$ corresponds to the
        fixed point.
  According to
       Pohlmeyer theorem \cite{POHLMEYER} if the propagator of the local scalar
       field is proportional to the propagator of massless free scalar field then the scalar field is free massless field.
       As a consequence of Pohlmeyer theorem the right hand side of the equation  (\ref{KLrho}) at $g_r = g_{r,fp} $ is equal to
       $ g^2_rZ^{-2}(\infty,...)   \frac{1}{16\pi^2} \theta(t)  $.
It means that
\begin{equation}
  lim_{t\rightarrow \infty} (g^2_r\rho_{\phi_{r2}\phi_{r3}, \phi^{*}_{r3}\phi^{*}_{r2}}(t)) = g^2_{r}Z^{-2}(\infty,...) \frac{1}{16\pi^2} \theta(t)
      \neq 0 \,.
    \label{133}
  \end{equation}
So we find the contradiction: the left hand side of the equation (\ref{KLrho}) vanishes in  the limit $t \rightarrow \infty$
while the right hand side of the equation (\ref{KLrho}) is different from zero in this limit. 

     %    At fixed point $g_{r,fp} $
     % the right hand side of the equation (\ref{KLrho}) is equal to zero while the left hand side of the equation (\ref{KLrho})
      % is equal to $g^2_{r}Z^{-2}(\infty,...) \frac{1}{16\pi^2}$. So we find the inconsistency of the equation  (\ref{KLrho})
       % t fixed point. It means that the fixed point behaviour ar $g_{r, fp} \neq 0$ is not possible \cite{NUOVOCIMENTO}.

   %    It should be noted that it is possible to prove the inconsistency of the WZ model without use of Pohlmeyer theorem.
   %    Namely as a consequence of the 
   %    commutation relations (\ref{Can3dr}) one can find that $\int_0^{\infty} dt \Delta \rho(\frac{t}{\Lambda^2}) = 0$.
   %        Hence for $\Delta \rho(\frac{t}{\Lambda^2}) \neq  0$ at some point $t= t_0$
   %        $\Delta \rho(\frac{t_0}{\Lambda^2}) <  0 $ that due to the equation  (\ref{KLrho}) contradicts to the
          %         non-negativity of the spectral density  $g^2_r\rho_{\phi_{r2}\phi_{r3}^{*} \phi^{*}_{r3}\phi^{*}_{r2}}(t)$.

% Another proof that fixed point regime is not realized is the following. At fixed point we have not also scale symmetry but
% also conformal symmetry. The conformal symmetry allows to fix the form of three point vertex function. In our case
% the three vertex     $ \Gamma_{3,F\phi\phi} = const$  and the propagator
% of the scalar field is proportional to the free scalar propagator. As a consequence the Schwinger equation for the propagator
% $D_{FF*}(p^2)$ is ultravioet divergent for   $ \Gamma_{3,F\phi\phi} = const  \neq 0$.  
 
   Note that the proof of the absence of fixed point in the WZ model is valid also for the general case.
   For instance,
 in full analogy with the previous consideration  one can prove that for the WZ model 
   with single chiral superfield   $\phi(x, \theta)$  and with the superpotential
  $W_3 =   \frac{m}{2}\phi^2(x, \theta) + \frac{g}{3}\phi^3(x,  \theta)$ fixed point behaviour is not possible  \cite{NUOVOCIMENTO}.

\section{Conclusions}

In this paper we  proved that  the supersymmetric WZ  model  is  internally contradictory.
Our proof is based on the following assumptions:

1. We assumed that the WZ model is the model with positively defined metric of states  - there are no ghost states in the
spectrum. As a
consequence of this assumption we used the KL inequality $0 \leq Z(\infty, g_r, \mu,  m_r) \leq 1$ for the wave function renormalization.

2. In the perturbation theory the Green's functions become ultraviolet finite by the introduction of the
$ \sum_{k=1}^{N}   \Delta Z_k\int d^2\theta d^2\bar{\theta}
\Phi^*_k(x, \theta, \bar{\theta})\Phi_k(x, \theta, \bar{\theta})$ counterterms. We assumed that this structure of the counterterms is
valid irrespective of the perturbation theory.

3. We assumed that commutation relations among generalized coordinates and
momenta  are valid in the renormalized field theory. 

Our  proof of the inconsistency of the finite wave function renormalization  $Z(\infty,...) \neq 0$
was based on the use of Polhlmeyer theorem while the proof for the case of infinite wave function
renormalization        $Z(\infty,...) =  0$            was based on the use of the commutation relations (\ref{Can3a} - \ref{Can3d}).
So our proof excludes  $Z(\infty,...) \neq  \pm \infty $.
%The finiteness of the wave function renormalization
%is direct consequence of the norm states positivity  $<n|n> > 0 $.
The case  $Z(\infty,...) =   \pm \infty $
corresponds to the  spectrum with negative norm states (so called Landau pole singularities).
  For $Z(\infty, m, \mu, g) = \pm \infty$  the Schwinger's equations   (\ref{Sch1})  for renormalized fields
  are not well defined and our proof based on the use of Schwinger's equations does not work.
  So   our results  confirm the hypothesis  that the spectrum of the WZ model contains the states with negative norm,  so called
  Landau poles.  The calculations in leading log approximation confirm 
this hypothesis.

I am indebted to the collaborators of the INR TH department for useful comments and discussions.

\end{document}